# INTERPRETING SELF-ORGANIZING MAPS THROUGH SPACE–TIME DATA MODELS[1]

By Huiyan Sang, Alan E. Gelfand, Chris Lennard,
Gabriele Hegerl and Bruce Hewitson

*Texas A&M University, Duke University, University of Cape Town,
University of Edinburgh and University of Cape Town*

Self-organizing maps (SOMs) are a technique that has been used with high-dimensional data vectors to develop an archetypal set of states (nodes) that span, in some sense, the high-dimensional space. Noteworthy applications include weather states as described by weather variables over a region and speech patterns as characterized by frequencies in time. The SOM approach is essentially a neural network model that implements a nonlinear projection from a high-dimensional input space to a low-dimensional array of neurons. In the process, it also becomes a clustering technique, assigning to any vector in the high-dimensional data space the node (neuron) to which it is closest (using, say, Euclidean distance) in the data space. The number of nodes is thus equal to the number of clusters. However, the primary use for the SOM is as a representation technique, that is, finding a set of nodes which representatively *span* the high-dimensional space. These nodes are typically displayed using maps to enable visualization of the continuum of the data space. The technique does not appear to have been discussed in the statistics literature so it is our intent here to bring it to the attention of the community. The technique is implemented algorithmically through a training set of vectors. However, through the introduction of stochasticity in the form of a space–time process model, we seek to illuminate and interpret its performance in the context of application to daily data collection. That is, the observed daily state vectors are viewed as a time series of multivariate process realizations which we try to understand under the dimension reduction achieved by the SOM procedure.

The application we focus on here is to synoptic climatology where the goal is to develop an array of atmospheric states to capture a collection of distinct circulation patterns. In particular, we have daily

Received November 2007; revised May 2008.
[1]Supported in part by NSF DEB 0516198.

weather data observed in the form of 11 variables measured for each of 77 grid cells yielding an $847 \times 1$ vector for each day. We have such daily vectors for a period of 31 years (11,315 days). Twelve SOM nodes have been obtained by the meteorologists to represent the space of these data vectors. Again, we try to enhance our understanding of dynamic SOM node behavior arising from this dataset.

**1. Introduction.** Self-organizing maps (SOMs) are a technique that has been used with high-dimensional data vectors to develop an archetypal set of states (nodes) that span, in some sense, the high-dimensional space. First developed by Kohonen (1995), the technique has subsequently found application to automatic speech recognition, analysis of electrical signals from the brain, data visualization and meteorology. See, for example, Ferrandez et al. (1997), Tamayo et al. (1999), Kaski (1997) and Crane and Hewitson (2003), respectively.

The SOM approach is essentially a neural network model that implements a nonlinear projection from a high-dimensional input space to a low-dimensional array of neurons. In the process, it also becomes a clustering technique, assigning to any vector in the high-dimensional data space the node/neuron (reference vector) to which it is closest (using, say, Euclidean distance) in the data space. The number of nodes is thus equal to the number of clusters. However, the primary use for the SOM is as a representation technique, that is, finding a set of nodes which representatively *span* the high-dimensional space. These nodes are typically displayed using maps to enable visualization of the continuum of the data space. Hence, the approach should not be viewed as an "optimal" clustering technique; in particular, in application it is expected to produce roughly equal cluster sizes.

A SOM algorithm is usually implemented in three stages. First, a specified number of nodes is selected and the values of the components for each node are initialized, typically selecting random values. Second, iterative training is performed where the nodes are adjusted in response to a set of training vectors so that the nodes approximately minimize an integrated distance criterion. The last stage of the SOM technique is visualization where each node's reference vector is projected in some fashion to a lower dimensional space and plotted as a map (perhaps several maps). Customary projection creates a set of neurons in two-dimensional space which arise as a deformation of a regular lattice. For a given training set, the frequency of occurrence of each node can be calculated as well as the average error at each node, the latter interpreted as a measure of coherence around the node. With regard to implementation, the number of nodes is arbitrary.

In any event, it is not our contribution here to criticize the SOM approach or to compare it with other clustering procedures. Rather, in practice, the procedure is implemented in a purely algorithmic manner, ignoring any spatial or temporal structure which may be anticipated in the training set. Our contribution is to attempt to incorporate structural dependence, through the introduction of stochasticity in the form of a space–time process model. As a result, we hope to illuminate and interpret the performance of the SOM procedure in the context of application to daily data collection. That is, the observed daily state vectors are viewed as a time series of multivariate spatial process realizations. Working with the original high-dimensional data renders formal modeling infeasible. Instead, we try to achieve this understanding through the dimension reduction achieved by the SOM procedure.

The application we focus on here is to synoptic climatology as introduced by Hewitson and Crane (2002), where the goal is to develop an array of atmospheric states to capture a collection of distinct circulations. There has been some literature on estimating synoptic states with the purpose of downscaling climate models. For example, Hughes, Guttorp and Charles (1999) and Bellone, Hughes and Guttorp (2000) propose nonhomogeneous hidden Markov models (NHMM) which relate precipitation occurrences and amounts at multiple rain gauge stations to broad-scale atmospheric circulation patterns. In particular, both papers assume that occurrences and precipitation amounts at each rain gauge are conditionally independent given the current synoptic weather state, that is, all the spatial dependence between rain gauges is induced by the synoptic weather state. See, also, the recent work of Vrac, Stein and Hayhoe (2007) in this regard.

In this paper we work with daily weather data observed in the form of 11 variables measured for each of 77 grid cells yielding an $847 \times 1$ vector for each day. We have such daily vectors for a period of 31 years (11,315 days). Twelve SOM nodes have been obtained by the meteorologists to represent the space of these data vectors. Fuller detail is provided in Section 3. We also note that a broader view of the use for a SOM in climatology is for inference at longer than daily time scales.

The format of the paper is as follows. In Section 2 we provide a brief review of the SOM theory and implementation. In Section 3 we detail the motivating dataset and some exploratory data analysis. In Section 4 we present a collection of models to investigate. Section 5 addresses model fitting issues, while Section 6 considers model selection and results. Section 7 offers some summary discussion.

**2. A review of SOM theory and implementation.** A self-organizing map (SOM) is a neural network model and algorithm that implements a nonlinear projection from a high-dimensional space of input vectors to a low-dimensional array of neurons. That is, input vectors are assigned to nodes (or neurons). Nodes have two positions, one in the high-dimensional space, say, in a subset of $\mathbb{R}^d$, one in the low-dimensional visualization space, typically taken to be a deformation of a regular lattice in two-dimensional space. For a given set of nodes $\{\mathbf{w}_1, \mathbf{w}_2, \ldots, \mathbf{w}_M\}$ in the high-dimensional space, an array index taking values in $\{j = 1, 2, \ldots, M\}$ is defined, for each $\mathbf{x} \in \mathbb{R}^d$, as $c(\mathbf{x}, \{\mathbf{w}_m\}) = j$ if $d(\mathbf{x}, \mathbf{w}_j) = \min_{1 \leq m \leq M} d(\mathbf{x}, \mathbf{w}_m)$ for some distance $d$ (usually Euclidean). The theoretical objective of the SOM is to minimize, over all choices of $\{\mathbf{w}_m, m = 1, 2, \ldots, M\}$, $\int g(d(\mathbf{x}, \mathbf{w}_{c(\mathbf{x}, \{\mathbf{w}_m\})})) p(\mathbf{x}) \, d\mathbf{x}$, where $g(\cdot)$ is a monotone function and $p(\mathbf{x})$ is the density function for the random input vectors in $\mathbb{R}^d$. Solution to this vector quantization problem is generally intractable. We note that if we confine $\mathbf{x}$ to a bounded rectangular subset of $\mathbb{R}^d$ and if $p(\mathbf{x})$ is assumed uniform over this subset, then, at the optimal $\{\mathbf{w}_m\}$, $c$ will be equally likely to take on each of its $M$ possible values. Hence, with a sample of $\mathbf{x}$'s from this uniform distribution, we expect equal numbers of the $\mathbf{x}$'s to be assigned to each of the index values, that is, to each of the nodes. A special version which seeks to minimize $\int \sum_m h(\mathbf{w}_{c(\mathbf{x}, \{\mathbf{w}_m\})}, \mathbf{w}_m) g(d(\mathbf{x}, \mathbf{w}_m)) p(\mathbf{x}) \, d\mathbf{x}$, where $h$ is a smoothing kernel, is customarily used. It lacks a closed form solution, but an approximate solution can be obtained iteratively using stochastic approximation [see Kohonen et al. (1996)] as we describe below.

We now offer a bit more detail on the nature of a SOM algorithm. In practice, the SOM procedure consists of three stages. Let $\{\mathbf{x}_i \in \mathbb{R}^d\}, i = 1, 2, \ldots, n$, denote the input training vectors. In our case, $d = 847$ reflecting the daily 847-element climate records from 1970 to 2000. SOMs seek to "optimally" place a specified number of nodes, $M$, again denoted by $\mathbf{w}_m \in \mathbb{R}^d, m = 1, 2, \ldots, M$. In the SOMs literature (and, as the default in the publicly available software package cited below) the suggested number of nodes is $5\sqrt{n}$. In our application, with $n \approx 10{,}000$, this would suggest roughly $M = 500$ nodes. However, climatologists categorize far fewer types of circulation patterns; for our South African data, they conclude that $M = 12$ is adequate. With 500 nodes, a two-dimensional representation offers the best prospects for visualization. However, with our 12 nodes describing 11 variables over an $11 \times 7$ grid, we can create more appropriate maps. For example, for each variable, we can provide 12 panels, each panel a map over the geographic space.

We describe two versions of the iterative training algorithm procedure of the SOM technique as follows:

- Initialization stage: given $M$, the node vectors are initialized with random values.
- Iterative training (Version 1):
  – At step $t$, randomly choose an input vector $\mathbf{x}^{(t)}$ from the training set $\{\mathbf{x}_i\}$ for $i = 1, \ldots, n$.
  – Compute the distance (e.g., Euclidean) between $\mathbf{x}^{(t)}$ and each of the node vectors $\mathbf{w}_m$. Identify the winning node $\mathbf{w}_{c(\mathbf{x}^{(t)})}$ whose node vector is closest to the input vector, that is, $\|\mathbf{w}_{c(\mathbf{x}^{(t)})} - \mathbf{x}^{(t)}\| \leq \|\mathbf{w}_m - \mathbf{x}^{(t)}\|$ for $m \in \{1, 2, \ldots, M\}$.

– Every node has its vector adjusted according to the following equation:
$$\mathbf{w}_m^{(t+1)} = \mathbf{w}_m^{(t)} + \alpha(t) K(m, c(\mathbf{x}^{(t)}))(\mathbf{x}^{(t)} - \mathbf{w}_m^{(t)}),$$
where $K(m, c(\mathbf{x}^{(t)}))$ is called the neighborhood function, and $\alpha(t)$ is called the learning rate, which is usually a decreasing function of step $t$. One example of $K(m, c(\mathbf{x}^{(t)}))$ is the Gaussian kernel $K(m, c(\mathbf{x}^{(t)})) = \exp\{-\|\mathbf{w}_m^{(t)} - \mathbf{w}_{c(\mathbf{x}^{(t)})}\|^2/2\sigma^2\}$. A simpler choice is a so-called "bubble" function, that is, a uniform over the neighborhood (Voronoi tessellation) of $\mathbf{w}_{c(\mathbf{x}^{(t)})}$, zero elsewhere.

Usually, the SOM training is performed in two phases. A relatively large initial learning rate is used in the first phase and a smaller learning rate is used in the second phase. This updating suggests that nodes close to the winner node, as well as the winner itself, update their vectors closer to $\mathbf{x}^{(t)}$ in the input data space. Vectors associated with far away output nodes do not change significantly.

– Repeat the above steps until the nodes converge. (Convergence is vaguely defined and is usually taken as the default in the software.)

- Iterative training (Version 2):
  – At step $t$, for each input vector $\mathbf{x}_i$ for $i = 1, \ldots, n$, compute the distance (e.g., Euclidean) between $\mathbf{x}_i$ and each of the node vectors $\mathbf{w}_m^{(t)}$. Identify the winning node $c(i)$ whose node vector is most similar to the input vector, that is, $\|\mathbf{w}_{c(i)}^{(t)} - \mathbf{x}_i\| \leq \|\mathbf{w}_m^{(t)} - \mathbf{x}_i\|$ for $m \in \{1, 2, \ldots, M\}$.
  – Every node has its vector adjusted according to the following equation:
  $$\mathbf{w}_m^{(t+1)} = \frac{\sum_{i=1:n} h_{m,c(i)}(t) \mathbf{x}_i}{\sum_{i=1:n} h_{m,c(i)}(t)},$$
  where $h_{m,c(i)}(t)$ is the neighborhood function around the winning node $c(i)$. One example is $h_{m,c(i)}(t) = \alpha(t) K(m, c(i))(t)$, where, again, $K(m, c(i))$ is the Gaussian neighborhood kernel $K(m, c(i)) = \exp\{-\|\mathbf{w}_m^{(t)} - \mathbf{w}_{c(i)}^{(t)}\|^2/2\sigma^2\}$. Here, $\alpha(t)$ is called the learning rate and is usually a decreasing function of step $t$.

  Again, the SOM training is performed in two phases. Again, a relatively large initial learning rate is used in the first phase and a smaller learning rate is used in the second phase. In this updating, the contribution (weight) of a particular training vector to each node only depends on the distance between the corresponding winning node of this training vector and each of the other nodes. $h_{m,c(i)}$ can be viewed as a smoothing function such that nodes close to the winner node as well as the winner itself update their vectors closer to the training vector in the input data space.

  – Repeat the above steps until the nodes converge.

- The final stage seeks to achieve visualization. When the number of nodes is large, visualization is most easily presented in two dimensions beginning with either a rectangular or hexagonal lattice of nodes. The iterative updating of the nodes eventually leads to a distortion of the lattice. (See Figure 4 and related discussion below.) An approach which is incorporated into the standard SOMs software (cited below) is the Sammon mapping scheme [Sammon (1969)]. Note that the goal here is "clustering" components of the vectors to achieve a two-dimensional representation, not clustering of the training vectors. In order to equalize the contributions of each of the components in the high-dimensional vectors with regard to classification, centering and scaling is recommended as pre-processing of the data. In our case this is certainly needed since the climate variables are measured over very different scales. Returning to the Sammon projection method, the basic idea is to arrange all the nodes on a 2-dimensional plane in such a way that the distances between the nodes in a 2-dimensional space resemble the distances in the original vector space as defined by some metric as faithfully as possible. Given the distance matrix $D$ with element $d_{i,j}$ being the distance between node $i$ and node $j$ according to some metric (e.g., Euclidean distance), our goal is to find $\mathbf{O}_m$ in $\mathbb{R}^2$ for each node $m$ for $m = 1, \ldots, M$ to minimize an error function $E$ defined by the following cost function: $E = \frac{1}{\sum \sum_{j>i} d_{i,j}} \sum_i \sum_{j>i} \frac{(d_{i,j} - \|\mathbf{O}_i - \mathbf{O}_j\|)^2}{d_{i,j}}$. Note that the $O$'s need a "center" to locate them. Also, the projections can be implemented at each iteration to see stability, hence convergence, as well as to assess interpretation.

A software package for implementing SOMs is available (<http://www.cis.hut.fi/research/som-research/>). For more detailed explanation of the SOM procedure, see the references and guidelines at this publicly available software website.

As a final comment on visualization, in our application below, with $M$ small and data associated with spatial locations, we can create maps in geographic space, as proposed above.

**3. The dataset and some exploratory data analysis.** The weather in a local region is conditional on the nature of the synoptic state of the atmosphere. Relating the synoptic scale characteristics to local scale responses requires the reduction of a large number of variables into a smaller set of data that still, in some sense, represent the original data. This goal motivated the use of the SOM technology. In this application we use daily multivariate weather data over a specified time period to produce generalized weather circulations. These are then easily visualized as an array of archetypal synoptic circulations that span the continuum of the data. In so doing, daily synoptic atmospheric data are categorized into a prescribed number of archetypal synoptic (circulation) modes characteristic of a specified time period. South African weather systems have been categorized into six to eight main "types" of circulation [Tyson and Preston-Whyte (2000)]. On testing various SOM sizes, a 12-node SOM was selected which was deemed to adequately represent all the expected synoptic types.

The SOM was trained using gridded ($2.5° \times 2.5°$), daily mean atmospheric fields constructed from six-hourly National Center for Environmental Prediction/National Center for Atmospheric Research (NCEP/NCAR) global reanalysis data [see Kalnay et al. (1996)]. Data were extracted for a domain with $11 \times 7$ grid cells over southern Africa whose latitudinal and longitudinal extent (25°S to 40°S; 10°E to 34.5°E) captures synoptic circulation patterns from the sub-tropics to the mid-latitudes. The following 11 variables were chosen as training data: mean sea level pressure, 500 hectopascal (hPa) geopotential height level, relative and specific humidities at the surface and at 700 hPa, daily maximum temperature at the surface, U- and V-wind components at the surface and at 700 hPa. Each of these variables was first standardized using the mean and the standard deviation of its corresponding $11 \times 7$ cell time series. These standardized variables were then used to create an 847-element vector ($11 \times 11 \times 7$) which described the daily atmospheric state. The time period from 1970 to 2000 was used, which resulted in 11,315 daily records that were used to train the SOM. (The eight extra days in the included leap years were not included in the analysis for computational purposes, but these would not significantly alter the results.) Each of the 11 climate variables was standardized separately to preserve the local gradients in each field.

The twelve resulting SOM nodes are labeled with their locations in two-dimensional space in Figure 1. This figure is intended to suggest that nodes near to each other are associated with somewhat similar synoptic states and that transition in SOM nodes is most likely to be to a neighboring node.

To clarify the visualization, the SOM of sea level pressure (SLP) is presented in Figure 2. It is used to assess the characteristic surface circulation associated with each node as it most clearly demonstrates the general synoptic circulation as well as associated regional weather patterns. Elaborating further, we briefly detail the features of the synoptic types captured by the 12 SOM nodes. South Africa is a semi-arid environment and can very generally be divided into summer and winter rainfall regions. Summer rainfall occurs over the interior of the country as a result of convective processes and winter rains over the south-western and southern coastal regions as a result of the passage of cold fronts. The SOM results show the majority (80%) of summer days map to nodes 5, 6, 8 and 9 and to a lesser degree nodes 3 and 12. These nodes are associated with characteristic summer circulation features. On the right-hand side of the SOM, a sub-tropical low pressure system is situated over the northern part of the domain which bring rainfall to the interior of the country. In nodes 8, 9 and 12 a high pressure system is located at relatively high latitudes to the south of the domain which pushes frontal systems southward and results in dry summers over the western parts of the country and introduces moisture to the eastern and central parts. In node 5 a linkage between the tropical low and mid-latitude circulation forms a tropical-temperate trough which results in rainfall over a large part of the interior of the country. The majority of winter days (over 70%) map to nodes on the left-hand side of the SOM (nodes 1, 4, 7, 10). To the south of the country, these nodes are associated with the west-east progression of mid-latitude cyclones (cold fronts) across the south of the country which bring rainfall to the south and south-western parts of the country and very cold temperatures, especially over the interior. Over the interior of the country, the sub-tropical low has moved northward and is replaced by a high pressure system which dominates the circulation resulting in cold, dry conditions. In nodes 7 and 10 a high pressure system brings cold, polar air into the country once the cold front has moved past. A typical winter synoptic sequence would be a progression from node 1 to node 4 to node 7 to node 10 over the period of about 2–3 days. Most spring days map to nodes 3, 10, 11, 12 and most autumn days map to nodes 1, 3 and 12. These nodes represent both summer and winter circulations expected in a transitional season.

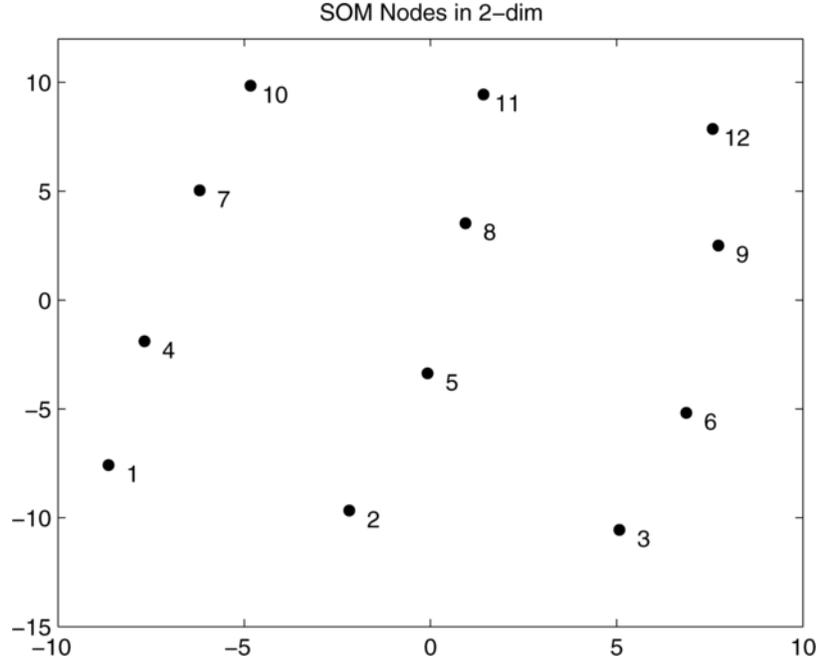

Fig. 1. *Projected locations of SOM nodes in 2-dimensional space.*

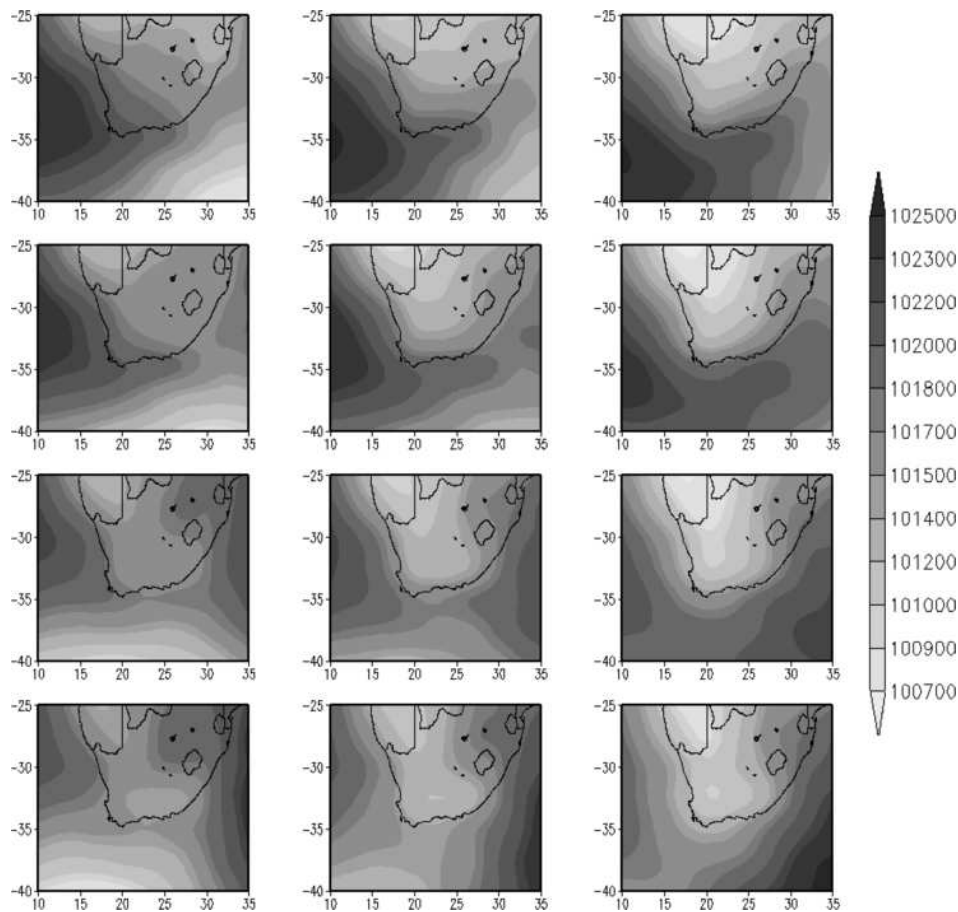

Fig. 2. *Sea level pressure (Pascals) associated with each node.*

Figure 3 is also presented in order to show a so-called 30 year climatology. It provides the departures, at grid cell level, from the 30 year mean (anomalies) of upper air specific humidity fields of each of the nodes. The anomaly map for specific humidity is more visually informative than the unadjusted map. Specific humidity in the upper atmosphere is used because it has been shown to be an important component for rainfall in the region in both summer and winter [Cavazos and Hewitson (2005)]. Over the interior of the country in the nodes representative of summer circulations, the high negative anomalies demonstrate the high moisture content in the atmosphere which (with other meteorological factors) result in wet summers. The high positive anomalies over the interior seen in the nodes associated with winter circulations indicate a lack of moisture in the atmosphere and dry winters. The opposite can be seen for the winter rainfall regions.

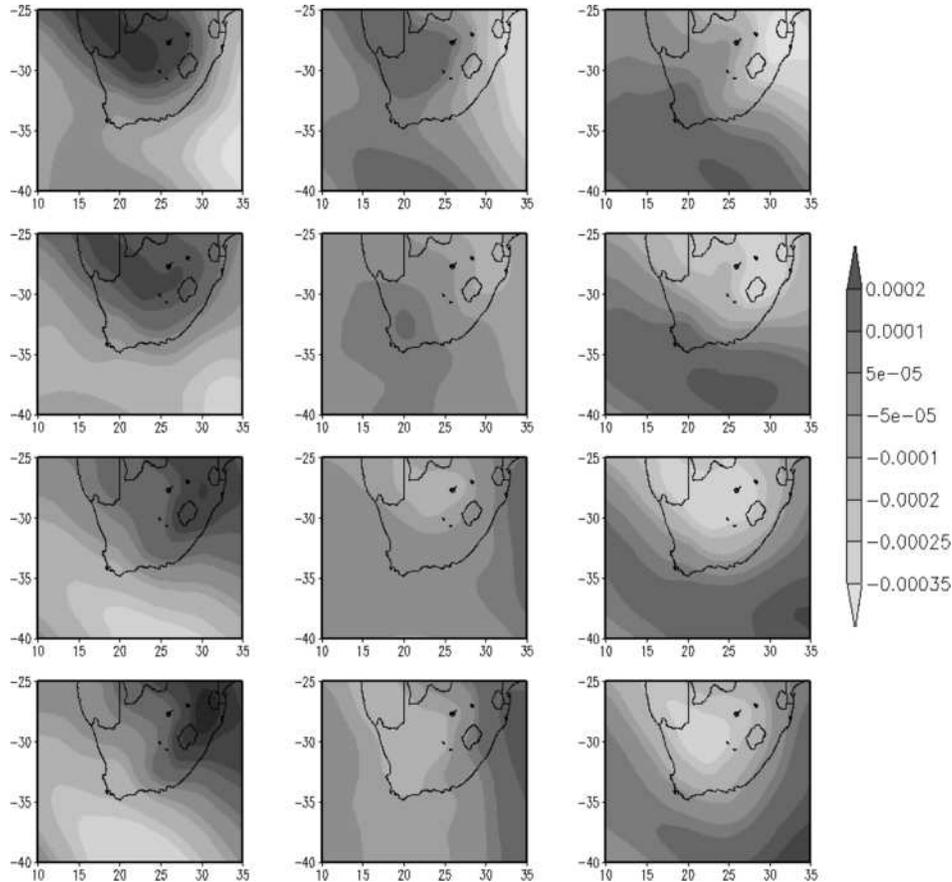

Fig. 3. *Upper air specific humidity anomaly field associated with each node.*

Undertaking some preliminary exploratory data analysis, a first investigation is to explore the frequencies of occurrence of each node, hence of the synoptic climate systems. Table 1 (left) provides a histogram showing the frequencies of daily observations mapped to each node over the entire study period, from which we observe fairly evenly distributed percentage frequency of occurrence for each synoptic node and no particular archetype is found dominant over the study period. This is an anticipated result of the SOM algorithm as clarified in Section 2. As a crude look to infer temporal behavior of the synoptic climate states, Table 1 (right) compares the histogram of frequency of occurrence during 1970 to 1979 along with that during 1990 to 1999. Some evidence of temporal shifting in the distribution of incidences over the study area is seen. For example, nodes 1, 2 and 3 occur more frequently during 1990 to 1999 compared with the 1970s. Table 2 (left) shows the frequency distribution of occurrence for each node in the summer season

TABLE 1
(Left) *Frequency of occurrence of each node over the entire study period, for example,* $935^{(4)}$ *indicates that the total number of occurrences of node 4 is 935;* (right) *Frequency of occurrence of each node during 1970–1979 and 1990–1999, for example,* $(278, 281)^{(5)}$ *indicates that the number of occurrence of node 5 during 1970–1979 is 278, and that number is 281 during 1990–1999*

| | | |
|---|---|---|
| $1070^{(10)}$ | $859^{(11)}$ | $1017^{(12)}$ |
| $989^{(7)}$ | $920^{(8)}$ | $910^{(9)}$ |
| $935^{(4)}$ | $869^{(5)}$ | $857^{(6)}$ |
| $1043^{(1)}$ | $776^{(2)}$ | $1070^{(3)}$ |
| $(340, 344)^{(10)}$ | $(320, 268)^{(11)}$ | $(336, 327)^{(12)}$ |
| $(291, 324)^{(7)}$ | $(308, 278)^{(8)}$ | $(341, 266)^{(9)}$ |
| $(291, 291)^{(4)}$ | $(278, 281)^{(5)}$ | $(290, 286)^{(6)}$ |
| $(283, 339)^{(1)}$ | $(242, 273)^{(2)}$ | $(330, 373)^{(3)}$ |

(December, January, February). It is clear that the climate archetype which corresponds to nodes 5, 6, 8 and 9 dominates during the summer period. As shown in Table 2 (right), the dominant climate archetypes transfer to another type of circulation in winter (June, July, August). Now, we see high frequency of SOM nodes 1, 4, 7 and 10. We may also use SOM arrays to examine short term (e.g., daily) temporal evolution of synoptic events. The frequencies of daily transitions from each node to other nodes are calculated and shown in Table 3, which reveals a somewhat clockwise cyclic evolution (with regard to Figure 1) of the weather systems. For example, SOM node group 9 displays preferential transition to group 6, while SOM node group 6 tends to most prefer transition to group 3. In Section 4 we elaborate this analysis by introducing formal time series modeling to interpret the SOM arrays.

TABLE 2
(Left) *Frequency of occurrence of each node during summer;* (right) *Frequency of occurrence of each node during winter*

| | | |
|---|---|---|
| $63^{(10)}$ | $149^{(11)}$ | $256^{(12)}$ |
| $86^{(7)}$ | $389^{(8)}$ | $416^{(9)}$ |
| $71^{(4)}$ | $402^{(5)}$ | $401^{(6)}$ |
| $56^{(1)}$ | $189^{(2)}$ | $312^{(3)}$ |
| $503^{(10)}$ | $198^{(11)}$ | $159^{(12)}$ |
| $463^{(7)}$ | $81^{(8)}$ | $56^{(9)}$ |
| $494^{(4)}$ | $59^{(5)}$ | $42^{(6)}$ |
| $530^{(1)}$ | $143^{(2)}$ | $124^{(3)}$ |

TABLE 3
*Empirical transition probabilities. Each 4 by 3 sub-table in the following 4 by 3 array shows a set of transition probabilities. The array and sub-tables are arranged in the same way as in Table 1*

| 0.244 | 0.180 | 0.107 | 0.051 | 0.148 | 0.278 | 0.007 | 0.025 | 0.246 |
|---|---|---|---|---|---|---|---|---|
| 0.144 | 0.064 | 0.032 | 0.036 | 0.097 | 0.134 | 0.008 | 0.029 | 0.204 |
| 0.091 | 0.035 | 0.018 | 0.024 | 0.063 | 0.058 | 0.002 | 0.026 | 0.184 |
| 0.043 | 0.024 | 0.019 | 0.015 | 0.027 | 0.069 | 0.009 | 0.045 | 0.216 |
| 0.166 | 0.104 | 0.045 | 0.040 | 0.071 | 0.114 | 0.002 | 0.012 | 0.076 |
| 0.227 | 0.080 | 0.022 | 0.043 | 0.201 | 0.170 | 0.005 | 0.042 | 0.211 |
| 0.190 | 0.038 | 0.013 | 0.035 | 0.136 | 0.086 | 0.016 | 0.120 | 0.254 |
| 0.079 | 0.022 | 0.013 | 0.033 | 0.038 | 0.034 | 0.015 | 0.066 | 0.180 |
| 0.163 | 0.071 | 0.020 | 0.045 | 0.077 | 0.059 | 0.008 | 0.028 | 0.064 |
| 0.207 | 0.066 | 0.007 | 0.061 | 0.183 | 0.086 | 0.013 | 0.056 | 0.074 |
| 0.224 | 0.034 | 0.003 | 0.089 | 0.166 | 0.072 | 0.021 | 0.135 | 0.183 |
| 0.174 | 0.016 | 0.014 | 0.087 | 0.047 | 0.028 | 0.049 | 0.133 | 0.236 |
| 0.212 | 0.043 | 0.007 | 0.106 | 0.088 | 0.026 | 0.050 | 0.061 | 0.041 |
| 0.186 | 0.039 | 0.007 | 0.072 | 0.107 | 0.017 | 0.018 | 0.040 | 0.018 |
| 0.170 | 0.012 | 0.004 | 0.086 | 0.115 | 0.012 | 0.030 | 0.081 | 0.039 |
| 0.283 | 0.025 | 0.013 | 0.228 | 0.116 | 0.028 | 0.093 | 0.259 | 0.269 |

## 4. SOM modeling.

4.1. *Dimension reduction.* The daily climate reference data consist of an $847 \times 1$ vector for each day within a 31-year period, which raises methodological and computational challenges when we attempt to interpret them in high dimension. In fact, since the $847 \times 1$ vector arises as an 11-dimensional vector at each of 77 grid units, it is clear that we are monitoring a multivariate space–time data process. We do not seek to model this process directly, a very challenging task to develop and, likely, infeasible to fit. Rather, we seek to understand this process in terms of the SOM nodes that have been created. We will take advantage of the dimension reduction provided by the SOM procedure to, instead, model the induced collection of two-dimensional locations across time. As we remarked earlier, the SOM algorithm ignores time and space in creating the nodes. By introducing a space–time process model, we seek to enhance behavioral interpretation for the set of SOM nodes. The result of the SOM algorithm yields, in our case, twelve nodes, each with an associated two-dimensional location (Figure 1). We now seek to *project* the 11,315 daily state vectors onto this space of locations. Many schemes are available to accomplish this; there is no "best" one. We choose to map daily high-dimensional reference data onto a 2-dimensional surface using high-dimensional pairwise distances along with the 2-dim coordinates of the SOM nodes. For each daily state vector, the Euclidean distances between it and each of the nodes are calculated in high-dimensional space. Then, for each daily vector, nodes are ranked by their distances to the vector. We next introduce a greedy space searching technique that maps each daily vector onto a 2-dimensional surface. To be specific, a 2-dimensional point within a finite bounded region is selected as the projection if the ranks of the Euclidean distances in two dimensions to each node using this point agree with the ranks between the vector and the nodes in high-dimensional space. Evidently, there may be no point in 2-dimensional space which satisfies this condition, so we seek agreement in ranking starting from the smallest distance. Also, for a given high-dimensional point, such mapping may yield multiple mapped positions that provide the same extent of agreement in terms of rank distance agreement. In that situation, we averaged the coordinates of the multiple positions to ensure the uniqueness of the mapped position. Such an algorithm is easy to code and easily handles 11,315 points in 847-dimensional space. In Section 4.2 we focus on modeling the 2-dimensional coordinates derived using this method. Other projection approaches utilizing alternative, possibly global, optimization criteria are certainly available, though they may be difficult to implement. However, the modeling approach we develop in Section 4.2 can be applied to the results of any projection strategy.

4.2. *Modeling specifications.*  The projection method described in Section 4.1 is performed on the daily referenced data from 1970–2000 to yield 2-dimensional mapped coordinates which are plotted in Figure 4 along with the coordinates of the SOM nodes. We can see that the two-dimensional space is naturally partitioned into 12 tessellations, each attached to a SOM node. Of course, Figure 4 gives no information regarding the temporal sequence of the points.

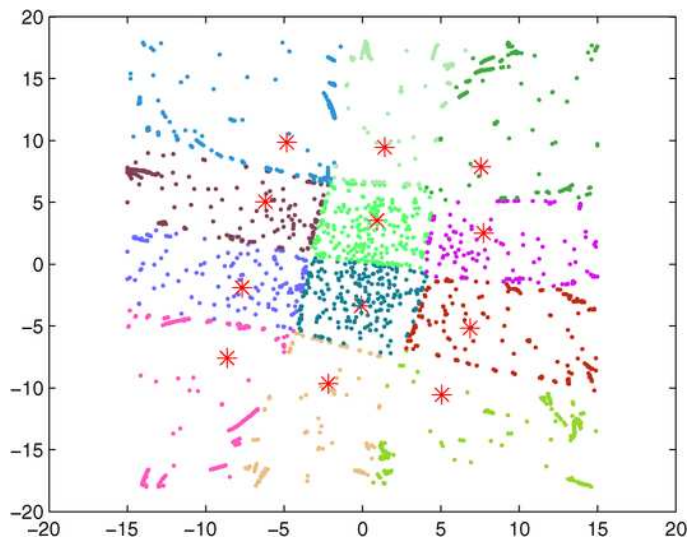

Fig. 4. *Mapped coordinates in the 2-dimensional space for each of the 11,315 days.*

However, let $\mathbf{s}_t = (x_t, y_t)'$ denote the coordinates in 2-dims for day $t$, $t = 1, \ldots, T$. Before beginning the time series analysis, it is natural to examine the autocorrelation in this bivariate time series. We ran a standard vector autoregression software package for lags 1 up to 50. The plot (not shown in the interest of space) finds an adjusted AIC value of 7.88 for the AR(1) model, 7.81 for the AR(2) model and reaches its minimum at 7.76 for the AR(24) model. So, while there may be some evidence of longer range dependence in the series, the relative decrease in the model choice criterion is very small; AR(1) models may be good enough. Moreover, with interest in studying transition probabilities, in the sequel we work exclusively with AR(1) specifications. Under 12 nodes this still yields 144 transition probabilities. For the AR(2), we arrive at 1728 transition probabilities, too many to estimate well and to display.

We start the analysis with a bivariate random walk Gaussian model

$$\mathbf{s}_{t+1} = \mathbf{s}_t + \boldsymbol{\varepsilon}_{t+1},$$

where $\boldsymbol{\varepsilon}_t$ follows a bivariate Gaussian distribution centered at 0 with a $2 \times 2$ covariance matrix $\boldsymbol{\Sigma}$. Under this model, the conditional Gaussian probability density functions of the coordinates at each time step are completely determined by the coordinates at the previous time step along with the covariance matrices. For us, the bivariate random walk model plays the role of a *straw man*. If the SOM nodes effectively capture synoptic weather states, there should be some structure to the daily transitions in the $\mathbf{s}_t$'s. In other words, the algorithm yielding the SOM nodes is applied to spatially-referenced explanatory climate variables observed over time and, therefore, we would expect behavior with a more mechanistic description than purely random movement of the daily states in two dimensions. In this regard, denote $\mathbf{Y}' = (\mathbf{s}_2, \ldots, \mathbf{s}_T)$, $\mathbf{X}' = (\mathbf{s}_1, \ldots, \mathbf{s}_{T-1})$. Then, the conditional maximum likelihood estimator (MLE) of $\boldsymbol{\Sigma}$ is

$$\boldsymbol{\Sigma} = (\mathbf{Y} - \mathbf{X})'(\mathbf{Y} - \mathbf{X})/(T-1).$$

A very general form of bivariate time series model is the following:

(1) $$\mathbf{s}_{t+1} = \mathbf{A}(\mathbf{s}_t, t)\mathbf{s}_t + \boldsymbol{\eta}(\mathbf{s}_t, t) + \boldsymbol{\varepsilon}_{t+1},$$

where $\mathbf{A}(\mathbf{s}_t, t)$ is a $2 \times 2$ unknown matrix containing autoregression coefficients that are allowed to vary in space and time, the values of which are specified by location and time at the preceding step. $\boldsymbol{\eta}(\mathbf{s}_t, t)$ enters (1) as an adjustment to the autoregressive component and can also be specified as a function of the preceding location and time. Again, error terms $\varepsilon_2, \ldots, \varepsilon_T$ are independently identically distributed $N(\mathbf{0}, \mathbf{\Sigma})$ representing unstructured noise or pure error in the model. The model in (1) provides a very flexible specification in the form of a locally affine transition model. In fact, it is also very challenging to fit. We are convinced that, even with more than 11,000 days of data, the data cannot support or identify such a general model; we can not achieve a well-behaved MCMC fitting algorithm. Hence, we turn to some model simplifications. We begin with specifications on $\mathbf{A}(\mathbf{s}_t, t)$:

- The first is a constant transformation matrix model $\mathbf{A}(\mathbf{s}_t, t) = \mathbf{A}$ yielding

$$(2) \qquad \mathbf{s}_{t+1} = \mathbf{A}\mathbf{s}_t + \varepsilon_{t+1}.$$

  This is a simple case of vector autoregressive (VAR) models, which have been widely used in multiple time series analysis [see, e.g., Sims (1972) and Enders (2003)].

- We next consider a spatially-varying transformation matrix. It is most convenient to assign a distinct transformation matrix to each of the tessellations induced by the SOM nodes as described above. Let $\mathbf{A}_l$ be the transformation matrix when $\mathbf{s}_t \in \mathbf{\Lambda}_l$, where $\mathbf{\Lambda}_l$ is tessellation $l$, for $l = 1, 2, \ldots, L$. Let $Z_l$ be a binary indicator, that is, $Z_l(\mathbf{s}_t) = 1$ if $\mathbf{s}_t \in \mathbf{\Lambda}_l$, and 0 otherwise. Then,

$$(3) \qquad \mathbf{A}(\mathbf{s}_t, t) = \sum_{l=1}^{L} \mathbf{A}_l Z_l(\mathbf{s}_t).$$

  This specification allows us to study regional change in the linear transformation.

- One of the important questions we seek to address in our climate study is whether there is a change in circulation patterns. That is, we assume the same collection of synoptic states continues to operate over time. However, temporal change would be manifested by a change in incidence rates of the states and thus would be modeled using time varying transition matrices. Let $\mathbf{B}_m$ be the transformation matrix when $t \in \mathbf{\Gamma}_m$, where $\mathbf{\Gamma}_m$ are disjoint time blocks, that is, $\bigcup_{m=1}^{M} \mathbf{\Gamma}_m = \{1, \ldots, T\}$. Let $V_m(t) = 1$ if $t \in \mathbf{\Gamma}_m$, and 0 otherwise. Now,

$$(4) \qquad \mathbf{A}(\mathbf{s}_t, t) = \sum_{m=1}^{M} \mathbf{B}_m V_m(t).$$

  Expression (4) enables us to study temporally varying linear transformation over suitable time scales, for example, months, quarters or years.

- A spatially and temporally varying structure can be extended from the specifications described above in the form

$$(5) \qquad \mathbf{A}(\mathbf{s}_t, t) = \sum_{l, m} \mathbf{D}_{m,l} Z_l(\mathbf{s}_t) V_m(t).$$

Special cases of (5) include *separable* forms in space and time, for example, $\mathbf{D}_{m,l} = \mathbf{B}_m \mathbf{A}_l$ or $\mathbf{D}_{m,l} = \mathbf{A}_l \mathbf{B}_m$. The former provides spatial transition followed by temporal transition, the latter vice versa.

The modeling in (3), (4) and (5) works at the aggregated spatial and temporal scale. Similarly, we could add spatially and temporally aggregated intercepts. In particular, we could introduce $\boldsymbol{\eta}(\mathbf{s}_t, t) = \boldsymbol{\eta}$, $\boldsymbol{\eta}(\mathbf{s}_t, t) = \sum_{l=1}^{L} \boldsymbol{\eta}_l Z_l(\mathbf{s}_t)$, or $\boldsymbol{\eta}(\mathbf{s}_t, t) = \sum_{m=1}^{M} \boldsymbol{\eta}_m V_m(t)$. However, we view the role of the $\boldsymbol{\eta}(\mathbf{s}_t, t)$'s as introducing point level refinement to aggregated level affine transformations. We do so by introducing a bivariate spatial Gaussian realization intended to provide spatially dependent adjustments to the linear transformation specification. The adjustment at time $t$ is $\boldsymbol{\eta}(\mathbf{s}_t)$, yielding the model

$$(6) \qquad \mathbf{s}_{t+1} = \mathbf{A}(\mathbf{s}_t, t)\mathbf{s}_t + \boldsymbol{\eta}(\mathbf{s}_t) + \boldsymbol{\varepsilon}_{t+1}.$$

We propose a coregionalization model for the bivariate Gaussian process realization in the spirit of Gelfand et al. (2004). Let $\mathbf{w}(s) = (w_1(\mathbf{s}), w_2(\mathbf{s}))'$, where $w_1(\mathbf{s})$ and $w_2(\mathbf{s})$ are uncorrelated spatial processes, each with zero mean and unit variance. Coregionalization constructs a bivariate spatial process by linear transformation of these two independent univariate processes, that is, $\boldsymbol{\eta}(s) = (\eta_1(\mathbf{s}), \eta_2(\mathbf{s}))' = \mathbf{Q}(w_1(\mathbf{s}), w_2(\mathbf{s}))'$, where $\mathbf{Q}$ is a $2 \times 2$ unknown coregionalization matrix and can be taken as lower triangular without loss of generality, that is, $\mathbf{Q} = \begin{pmatrix} q_{11} & 0 \\ q_{12} & q_{22} \end{pmatrix}$. An unusual aspect of the employment of this bivariate specification is that it provides a spatial surface to smooth all locations in the region while the observations are, in fact, a time series of locations. In other words, this bivariate spatial process is created for observed locations at multiple time points rather than multiple locations observed at the same time. The process realization reflects the spatial variation unexplained by the autoregressive component, regardless of the specific times of the observations.

The model in (6) is now completely specified. However, recall that we work with 11,315 days, hence 11,315 locations in total. The joint distribution of the collection of 11,315 $\boldsymbol{\eta}(\mathbf{s})$'s introduces an $11{,}315 \times 11{,}315$ covariance matrix. To handle this dimension, we employ a version of the predictive process model described in Banerjee et al. (2008). Briefly, we consider a set of "knots" $\mathcal{S}^* = \{\mathbf{s}_1^*, \ldots, \mathbf{s}_m^*\}$, which forms a subset of the study region in 2-dimensional space. The bivariate Gaussian process above would yield $\mathbf{w}^* = [w(\mathbf{s}_i^*)]_{i=1}^{m} \sim MVN(\mathbf{0}, C^*(\boldsymbol{\theta}))$ as its realizations over $\mathcal{S}^*$, where $C^*(\boldsymbol{\theta}) = [C(\mathbf{s}_i^*, \mathbf{s}_j^*; \boldsymbol{\theta})]_{i,j=1}^{m}$ is the corresponding $m \times m$ covariance matrix. The predictive process model is defined as

$$(7) \qquad \tilde{w}(\mathbf{s}) = E[w(\mathbf{s})|\mathbf{w}^*] = \mathbf{c}^T(\mathbf{s}; \boldsymbol{\theta}) C^{*-1}(\boldsymbol{\theta}) \mathbf{w}^*,$$

where $\mathbf{c}(\mathbf{s}; \boldsymbol{\theta}) = [C(\mathbf{s}, \mathbf{s}_j^*; \boldsymbol{\theta})]_{j=1}^{m}$. In fact, $\tilde{w}(\mathbf{s})$ is a Gaussian process with covariance function $\tilde{C}(\mathbf{s}, \mathbf{s}'; \boldsymbol{\theta}) = \mathbf{c}^{*T}(\mathbf{s}; \boldsymbol{\theta}) C^{*-1}(\boldsymbol{\theta}) \mathbf{c}^*(\mathbf{s}', \boldsymbol{\theta})$, where $\mathbf{c}^*(\mathbf{s}; \boldsymbol{\theta}) =$ $[C(\mathbf{s}_0, \mathbf{s}_j^*; \boldsymbol{\theta})]_{j=1}^m$. The realization of $\tilde{w}(\mathbf{s})$ on any collection of sites is the interpolated predictions conditional upon the realization of $w(\mathbf{s})$ over $\mathcal{S}^*$. To work with this process, we only need to work with $\mathbf{w}_1^*$, $\mathbf{w}_2^*$ and the associated pair of $m \times m$ correlation matrices.

**5. Model fitting issues.** VAR models are well discussed in the literature [see, e.g., Lütkepohl (1993) and Zivot and Wang (2006)]. Analysis within the Bayesian paradigm is presented in, for example, Sims and Zha (1998), Sun and Ni (2004). We employ MCMC to fit the various submodels of (6) described in the previous section. In fact, we first discuss the computational issues in fitting the proposed models without spatial adjustment. Then we turn to issues in fitting the models with such adjustment.

5.1. *Fitting models without spatial adjustment.* For those proposed models without spatial adjustment, we illustrate the MCMC fitting procedure for model $\mathbf{s}_{t+1} = \sum_{l=1}^L \mathbf{A}_l Z_l(\mathbf{s}_t) \mathbf{s}_t + \boldsymbol{\varepsilon}_{t+1}$.

Denote $\mathbf{Z}(\mathbf{s}_t) = (Z_1(\mathbf{s}_t), \ldots, Z_L(\mathbf{s}_t))$, $\mathbf{x}_t = (\mathbf{Z}(\mathbf{s}_t) \otimes \mathbf{s}_t')$, $\mathbf{Y}' = (\mathbf{s}_2, \ldots, \mathbf{s}_T)$, $\mathbf{X}' = (\mathbf{x}_1, \ldots, \mathbf{x}_{T-1})$, $\boldsymbol{\varepsilon}' = (\boldsymbol{\varepsilon}_2, \ldots, \boldsymbol{\varepsilon}_T)$, $\boldsymbol{\Phi}' = (\mathbf{A}_1', \ldots, \mathbf{A}_L')$. Here $\mathbf{Y}$ and $\boldsymbol{\varepsilon}$ are $(T-1) \times 2$ matrices, $\boldsymbol{\Phi}$ is a $2L \times 2$ matrix of unknown transformation parameters, $\mathbf{x}_t$ is a $1 \times 2L$ row vector and $\mathbf{X}$ is a $(T-1) \times 2L$ matrix of observations. Then the VAR model can be written as

$$\mathbf{Y} = \mathbf{X}\boldsymbol{\Phi} + \boldsymbol{\varepsilon}.$$

The MLE's of $\boldsymbol{\Phi}$ and $\boldsymbol{\Sigma}$ are obtained by maximizing

$$L(\boldsymbol{\Phi}, \boldsymbol{\Sigma}) = \frac{1}{|\boldsymbol{\Sigma}|^{(T-1)/2}} \exp\left\{-\frac{1}{2} \sum_{t=1}^{T-1} (\mathbf{s}_{t+1} - \mathbf{x}_t \boldsymbol{\Phi}) \boldsymbol{\Sigma}^{-1} (\mathbf{s}_{t+1} - \mathbf{x}_t \boldsymbol{\Phi})'\right\}$$

$$= \frac{1}{|\boldsymbol{\Sigma}|^{(T-1)/2}} \operatorname{etr}\left\{-\frac{1}{2} (\mathbf{Y} - \mathbf{X}\boldsymbol{\Phi}) \boldsymbol{\Sigma}^{-1} (\mathbf{Y} - \mathbf{X}\boldsymbol{\Phi})'\right\},$$

where $\operatorname{etr}(\cdot) = \exp(\operatorname{trace}(\cdot))$. We obtain MLE's of $\boldsymbol{\Phi}$ and $\boldsymbol{\Sigma}$ as $\hat{\boldsymbol{\Phi}}_M = (\mathbf{X}'\mathbf{X})^{-1}\mathbf{X}'\mathbf{Y}$ and $\hat{\boldsymbol{\Sigma}}_M = (\mathbf{Y} - \mathbf{X}\hat{\boldsymbol{\Phi}}_M)'(\mathbf{Y} - \mathbf{X}\hat{\boldsymbol{\Phi}}_M)/(T-1)$.

Bayesian model fitting is completed by assigning prior distributions on the unknown parameters of interest. Denote $\boldsymbol{\phi} = \operatorname{vec}(\boldsymbol{\Phi})$, that is, the vector obtained by concatenating the entries in $\boldsymbol{\Phi}$. We assign $\boldsymbol{\phi}$ with a flat prior. We consider a noninformative Jeffreys prior for $\boldsymbol{\Sigma}$, which, in our case, is $\pi(\boldsymbol{\Sigma}) \propto \frac{1}{|\boldsymbol{\Sigma}|^{3/2}}$.

Given $\boldsymbol{\Sigma}$, we can directly sample $\boldsymbol{\phi}$ from its conditional distribution given by

$$\pi(\boldsymbol{\phi}|\boldsymbol{\Sigma}, \mathbf{Y}) \sim MVN(\operatorname{vec}(\hat{\boldsymbol{\Phi}}_M), \boldsymbol{\Sigma} \otimes (\mathbf{X}'\mathbf{X})^{-1}).$$

Conditional on $\boldsymbol{\phi}$, $\boldsymbol{\Sigma}$ is updated using an Inverse Wishart $((T-1)\hat{\boldsymbol{\Sigma}}(\hat{\boldsymbol{\Phi}}_M), T - 2L)$.

5.2. *Fitting models with spatial adjustment.* For the models with spatial adjustment, for convenience, we adopt an exponential correlation function for each of the two parent processes, hence bringing in two decay parameters $\theta_1$ and $\theta_2$.[1] A uniform prior is assigned for each of $\theta_1$ and $\theta_2$ and updated using Metropolis steps. For the coregionalization matrix $\mathbf{Q}$, we assign truncated normal priors with positive support for the diagonal entries, and a normal prior for the off-diagonal entry. The entries in $\mathbf{Q}$ are updated from normal distributions conditional on the other parameters. Samples of $\mathbf{W}_1^{*(b)}$ and $\mathbf{W}_2^{*(b)}$ are generated in blocks from their multivariate normal posterior distributions, which in turn yield samples of $\tilde{w}(\mathbf{s})^{(b)} = \mathbf{c}^T(\mathbf{s}; \boldsymbol{\theta}^{(b)}) C^{*-1}(\boldsymbol{\theta}^{(b)}) \mathbf{w}^{*(b)}$. Substantial gains in computational efficiency are achieved by working with $\mathbf{W}^*$ at a relatively small number of knots.

Each of the proposed models enables one day ahead prediction, that is, the posterior predictive distribution of location at time $t+1$ given location, say, $\tilde{\mathbf{s}}$ at time $t$. This is implemented by composition; a posterior draw of the parameters in whatever version of (6) we fit, setting $\mathbf{s}_t = \tilde{\mathbf{s}}$, enables a predictive draw for the location at time $t+1$. Posterior samples, $\tilde{\mathbf{s}}^{(b)}$ enable a density estimate for the transition distribution at any time and given any location. In addition, these models enable inference about the "transition distance," $\|\mathbf{s}_{t+1} - \mathbf{s}_t\|$, in 2-dimensional space.

In fact, again using posterior predictive samples, these models allow us to induce inference for categorical analysis at tessellation level. The probability of transitioning from $\tilde{\mathbf{s}}$ to tessellation $l$ can be straightforwardly estimated as well as the $12 \times 12$ transition matrix from SOM node to SOM node (we omit details). This model-based estimate can be compared with the empirical estimate presented in Section 3 (Figure 3). Evidently, we can learn about the movement of the daily state vectors at any spatial scale in 2-dimensions. Working at the scale of the tessellations enables us to inform about circulation among synoptic weather states defined by the SOM nodes.

**6. Model comparison and model results.** Given the various possible model specifications detailed in Section 3, our first analysis goal would appear to be model comparison. We consider three criteria. First, we compare models in terms of one step ahead prediction performance. For each observation $\mathbf{s}_t$ at $t$, samples of $\mathbf{s}_{t+1}$ are drawn from the posterior predictive distribution as described above. Posterior means are adopted as the point estimates of the predicted positions for each of $t = 2$ to $t = T$. The root mean square predictive errors (*RMSPE*) are computed to assess the predictive performance for

---

[1] We would not anticipate sensitivity in the bivariate predicted $\boldsymbol{\eta}$ surfaces to the choice of correlation function.

each model

$$\widehat{RMSPE} = \sqrt{\frac{1}{T-1} \sum_{i=2}^{T} \|\hat{\mathbf{s}}_t - \mathbf{s}_t\|^2}, \tag{8}$$

where $\hat{\mathbf{s}}_t$ is the mean of the posterior samples $\{\hat{\mathbf{s}}_t^{(b)}\}$ for $b = 1, 2, \ldots, B$.

A second comparison among models is to study the proportion of times the true locations lie in their associated 95% predictive interval under each model. This coverage proportion is denoted by $\hat{r}$. A third model selection criterion which is easily calculated from the posterior samples is the deviance information criterion (*DIC*) [Spiegelhalter et al. (2002)]. *DIC* is a generalization of the AIC and BIC criteria; smaller values of *DIC* correspond to preferred models.

Table 4 summarizes the *RMSPE*, *DIC* score, and $\hat{r}$ for a collection of models as indicated. (When time is included it is either blocked quarterly or annually.) Based upon the model fitting to more than 11,000 data points and using a large number of posterior samples (10,000), we are comfortable with the number of significant digits provided. First, all of the proposed autoregressive models are apparently superior to the random walk model in terms of predictive performance and *DIC* scores. Second, disappointingly, the models including spatial adjustment show no evidence of improving performance on data fitting and prediction; there appears to be little spatial dependence left in the autoregression residuals. Further disappointment emerges in the similarity of performance of these models; we can do better than a random walk model, but can not find any interesting spatial or temporal structure.

We offer several thoughts in this regard. Perhaps the projection to a two-dimensional space which yields our bivariate time series has removed the interesting structure. In particular, the spatially varying covariate information associated with the 11 climate variables was used to create the projected locations in two dimensions; it is not available to explain the bivariate time series. Moreover, the spatial dependence, that is, induced in the two-dimensional space may have little to do with the spatial structure in the geographic space of the original 11 space–time processes. Finally, climatologists would assert that the SOM which was created is not intended for short term weather prediction; in capturing climate states, the SOM might be more appropriate for assessing regional climate change over a longer temporal scale (see Section 6). So, while our modeling goal here was to learn about spatial and temporal structure in the created SOM (and, what follows below indicates that there is still a story to tell), to learn about the space time structure in the original daily data a different dimension-reduction strategy might be more appropriate.

TABLE 4
*Performance of several VAR models using root mean square predictive errors (RMSPE), the deviance information criterion (DIC) and the empirical coverage probability $\hat{r}$. $\mathbf{A}(tessellation)$ denotes regionally varying $\mathbf{A}$ as described in (3), $\mathbf{A}(year)$ denotes annually varying $\mathbf{A}$ as described in (4), $\boldsymbol{\eta}(quarter)$ denotes quarterly varying $\mathbf{A}$ which is constant as year within each quarter. $\boldsymbol{\eta}(quarter^*)$ denotes quarterly varying $\mathbf{A}$ which is also changing as year, $\mathbf{A}(tessellation, quarter)$ denotes $\mathbf{A}$ as described in (5) and $\boldsymbol{\eta}(spatial)$ denotes spatially varying adjustment as described in (6)*

|  | Model specification | RMSPE | DIC ($\times 10^5$) | $\hat{r}$ |
|---|---|---|---|---|
| Model 0: | $\mathbf{s}_{t+1} = \mathbf{s}_t + \boldsymbol{\varepsilon}$ | 11.419 | 1.6403 | 0.929 |
| Model 1: | $\mathbf{s}_{t+1} = \mathbf{A}\mathbf{s}_t + \boldsymbol{\varepsilon}$ | 9.226 | 1.5353 | 0.944 |
| Model 2: | $\mathbf{s}_{t+1} = \mathbf{A}(tessellation)\mathbf{s}_t + \boldsymbol{\varepsilon}$ | 9.125 | 1.5317 | 0.943 |
| Model 3: | $\mathbf{s}_{t+1} = \mathbf{A}(tessellation)\mathbf{s}_t + \boldsymbol{\eta}(quarter) + \boldsymbol{\varepsilon}$ | 8.943 | 1.5238 | 0.939 |
| Model 4: | $\mathbf{s}_{t+1} = \mathbf{A}(quarter)\mathbf{s}_t + \boldsymbol{\varepsilon}$ | 9.215 | 1.5337 | 0.942 |
| Model 5: | $\mathbf{s}_{t+1} = \mathbf{A}(quarter)\mathbf{s}_t + \boldsymbol{\eta} + \boldsymbol{\varepsilon}$ | 9.205 | 1.5337 | 0.941 |
| Model 6: | $\mathbf{s}_{t+1} = \mathbf{A}(quarter)\mathbf{s}_t + \boldsymbol{\eta}(year) + \boldsymbol{\varepsilon}$ | 9.190 | 1.5297 | 0.947 |
| Model 7: | $\mathbf{s}_{t+1} = \mathbf{A}(quarter^*)\mathbf{s}_t + \boldsymbol{\varepsilon}$ | 9.088 | 1.5378 | 0.948 |
| Model 8: | $\mathbf{s}_{t+1} = \mathbf{A}(year)\mathbf{s}_t + \boldsymbol{\varepsilon}$ | 9.224 | 1.5361 | 0.944 |
| Model 9: | $\mathbf{s}_{t+1} = \mathbf{A}(tessellation, year)\mathbf{s}_t + \boldsymbol{\varepsilon}$ | 8.777 | 1.5407 | 0.956 |
| Model 10: | $\mathbf{s}_{t+1} = \mathbf{A}(tessellation, quarter)\mathbf{s}_t + \boldsymbol{\varepsilon}$ | 8.920 | 1.5247 | 0.940 |
| Model 11: | $\mathbf{s}_{t+1} = \mathbf{A}\mathbf{s}_t + \boldsymbol{\eta}(spatial) + \boldsymbol{\varepsilon}$ | 9.214 | 1.5421 | 0.934 |

In any event, Model 9, which has transformation matrix $A$ specified as tessellation and year, has the lowest prediction error in terms of *RMSPE*.[2] In addition, the $\hat{r}$ for Model 9 is quite close to 0.95, as desired. As a result, we summarize results based on Model 9. Table 5 provides the posterior means for several parameters of interest and the corresponding 95% credible intervals under Model 9. We notice that estimates for the elements in $\boldsymbol{\Sigma}$ take large values, suggesting substantial unexplained variances in the SOM array. Again, this reflects our lack of covariate information but also comments upon the utility of SOMs for one-step ahead prediction of weather states.

Following the discussion in Section 5, we provide some illustrations of the induced categorical analysis at the scale of the tessellations. Figure 5 shows the histograms of transition distance for each of the 12 tessellations in the year 2000, which suggests possible regional heterogeneity in the distribution of transition distance. In fact, synoptic weather in South Africa may display node-specific magnitude in volatility. For instance, SOM node group 2 which is expected in a transitional season, on average, tends to have higher transition distances than SOM node group 5 associated with characteristic summer circulation features. And this phenomenon might reveal high climate volatility associated with SOM node group 2 and relatively low climate volatility associated with SOM node group 5. Model 9 enables us to make inference of the transition matrices, as well as the corresponding estimated errors year by year. The estimated transition matrix is shown in Table 6 for the year 2000, which again supports the findings on the clockwise cyclic evolution of the weather systems. The trajectories of transition probabilities can be aligned in each row for 31 consecutive years, from which we can examine the temporal behavior of the transition probabilities in synoptic climate states. As an illustration, Figure 6 plots the trajectories of three selected transition probabilities. For the general synoptic state associated with SOM node 1, the persistence probability reached above 0.35 in 1986 and then dropped below 0.15 in the subsequent year. The transition probability from SOM node group 11 to SOM node group 12 roughly remained at a stationary level during the 1970s, then dramatically fell below 0.15 in 1983. It appears to be evolving with a more volatile trajectory since the late 1980's, finally reaching a peak which is above 0.35 in 2000.

---

[2]We might speculate that annual blocking captures an El Nino effect.

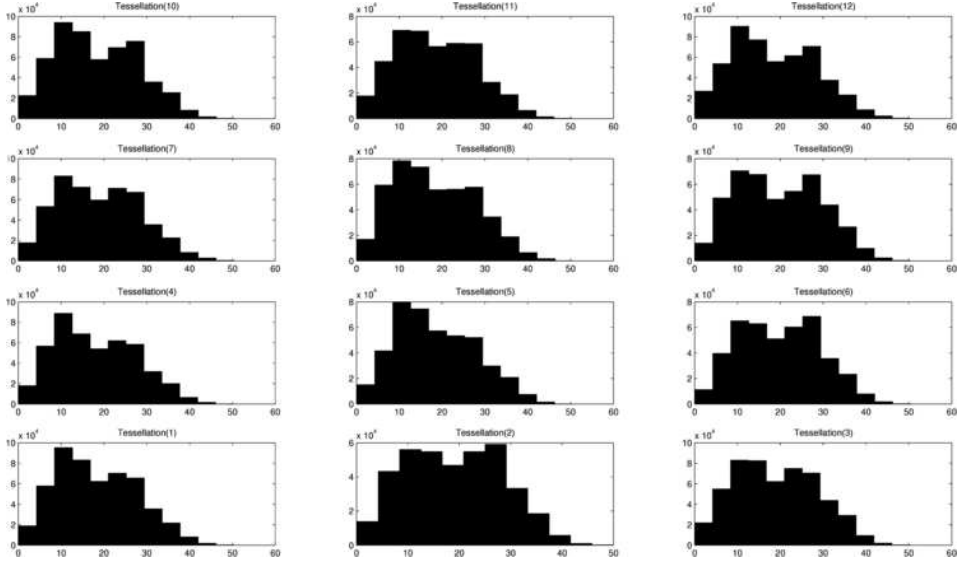

Fig. 5. *Histograms of transition distances for each SOM group.*

TABLE 5
*Posterior means and 95% credible intervals of $\mathbf{A}_{tessellation1, year2000}$ and $\mathbf{\Sigma}$*

|            | Mean    | 0.025%  | 0.975%  | SE    |
|------------|---------|---------|---------|-------|
| $a_{11}$   | 0.604   | 0.151   | 1.033   | 0.237 |
| $a_{12}$   | 0.168   | −0.220  | 0.618   | 0.226 |
| $a_{21}$   | −0.199  | −0.739  | 0.409   | 0.286 |
| $a_{22}$   | −0.028  | −0.678  | 0.503   | 0.289 |
| $\Sigma_{11}$ | 40.394 | 39.212 | 41.457 | 0.566 |
| $\Sigma_{12}$ | 17.572 | 16.544 | 18.648 | 0.531 |
| $\Sigma_{22}$ | 68.441 | 66.429 | 70.297 | 0.961 |

**7. Discussion.** The use of SOMs has made considerable inroads in the meteorology community with regard to developing synoptic weather states to describe the collection of available weather patterns across a region. Since the SOM technology represents high-dimensional vectors in two-dimensional space, we considered vector AR models to try to better understand the temporal evolution of SOM nodes. We have demonstrated that, while these SOMs may adequately span the high-dimensional space of daily weather data vectors, they reveal little interesting spatial or temporal structure with regard to forecasting weather states.

Finally, we offer a potentially useful remark. The inability of the SOM to predict short term temporal evolution of these states does not imply that the SOM will not be useful for the projection of future climate. If we assume that the SOM nodes describe regional weather well and that the same weather states continue to operate in the future, we may be able to forecast climate change in the form of a less uniform incidence of the different states than we currently see (Tables 1 and 2).

Table 6

*Estimated mean of the transition matrix in 2000. Each 4 by 3 sub-table in the following 4 by 3 array shows a set of transition probabilities. The array and sub-tables are arranged in the same way as in Table 1*

| 0.144 | 0.169 | 0.150 | 0.042 | 0.205 | 0.458 | 0.003 | 0.022 | 0.162 |
|-------|-------|-------|-------|-------|-------|-------|-------|-------|
| 0.081 | 0.100 | 0.081 | 0.017 | 0.060 | 0.104 | 0.003 | 0.036 | 0.231 |
| 0.059 | 0.074 | 0.048 | 0.011 | 0.035 | 0.040 | 0.005 | 0.048 | 0.249 |
| 0.030 | 0.037 | 0.026 | 0.004 | 0.011 | 0.013 | 0.004 | 0.041 | 0.194 |
| 0.177 | 0.108 | 0.040 | 0.027 | 0.085 | 0.159 | 0.008 | 0.035 | 0.106 |
| 0.161 | 0.087 | 0.025 | 0.031 | 0.100 | 0.146 | 0.012 | 0.063 | 0.161 |
| 0.148 | 0.073 | 0.016 | 0.035 | 0.107 | 0.119 | 0.017 | 0.091 | 0.194 |
| 0.103 | 0.049 | 0.012 | 0.029 | 0.077 | 0.085 | 0.019 | 0.094 | 0.200 |
| 0.139 | 0.106 | 0.041 | 0.057 | 0.128 | 0.130 | 0.008 | 0.027 | 0.060 |
| 0.132 | 0.095 | 0.031 | 0.055 | 0.119 | 0.094 | 0.013 | 0.053 | 0.112 |
| 0.133 | 0.092 | 0.024 | 0.057 | 0.115 | 0.070 | 0.019 | 0.086 | 0.177 |
| 0.107 | 0.076 | 0.023 | 0.043 | 0.081 | 0.052 | 0.024 | 0.124 | 0.297 |
| 0.187 | 0.064 | 0.013 | 0.089 | 0.082 | 0.041 | 0.014 | 0.024 | 0.025 |
| 0.197 | 0.055 | 0.008 | 0.116 | 0.091 | 0.033 | 0.028 | 0.054 | 0.053 |
| 0.203 | 0.050 | 0.006 | 0.145 | 0.097 | 0.026 | 0.049 | 0.103 | 0.098 |
| 0.168 | 0.042 | 0.005 | 0.165 | 0.091 | 0.024 | 0.085 | 0.227 | 0.240 |

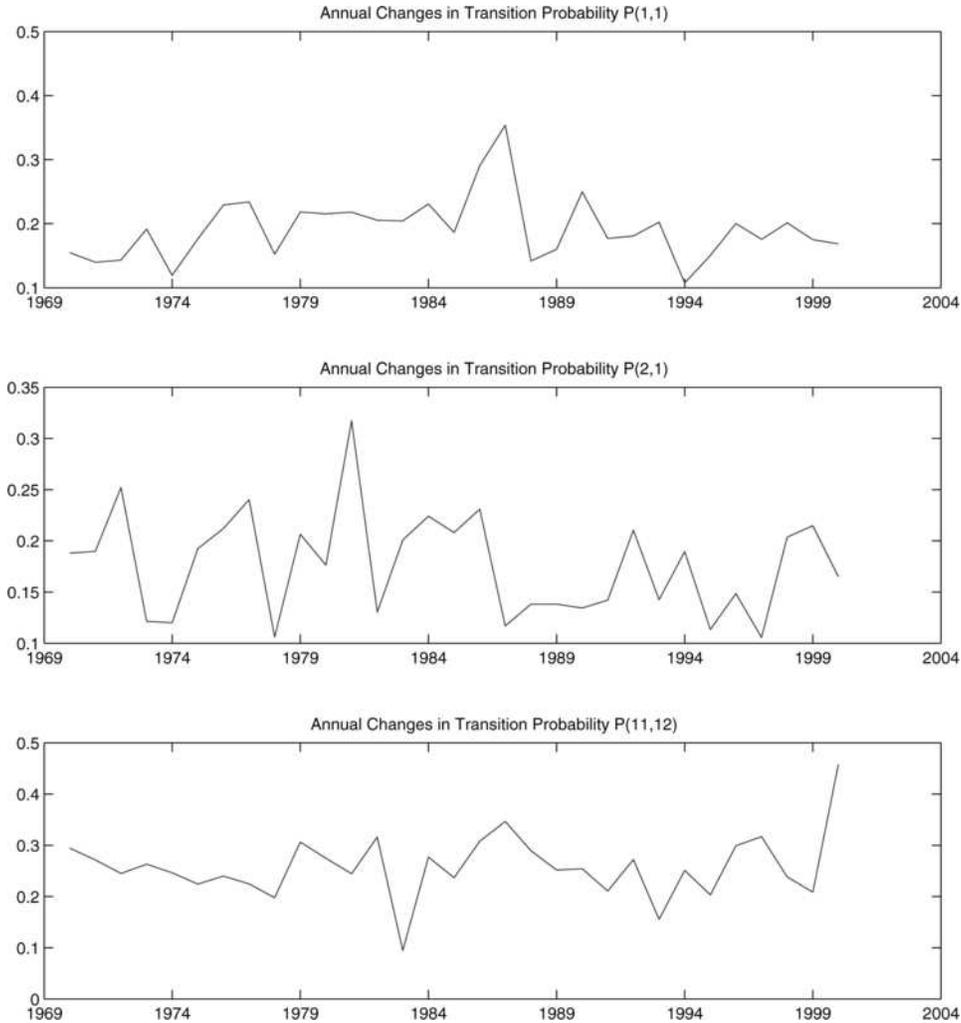

Fig. 6. *Transition probabilities from 1970–2000.*

**Acknowledgments.** The authors thank John Silander, Jr. and Andrew Latimer for useful conversations in the development of this work and the Editor for invaluable comments to improve the manuscript.

H. Sang
Department of Statistics
Texas A&M University
College Station, Texas 77843
USA
E-mail: huiyan@stat.tamu.edu

A. E. Gelfand
Department of Statistical Science
Duke University
Durham, North Carolina 27708
USA
E-mail: alan@stat.duke.edu

C. Lennard
B. Hewitson
The Climate Systems Analysis Group
University of Cape Town
Cape Town
South Africa
E-mail: lennard.chris@gmail.com
       hewitson@egs.uct.ac.za

G. Hegerl
School of Geosciences
University of Edinburgh
United Kingdom
E-mail: Gabi.Hegerl@ed.ac.uk